\documentclass[]{jpsj2}
\usepackage{color}
\newcommand{\vectr}[1]{\mbox{\boldmath $#1$}}
\begin{document}
%\setlength{\baselinestretch}{1.5}
% 2002/12/16
%\usepackage[mtbold]{mathtime}
\def\runtitle{Instructions for the Preparation of Manuscript}
\def\runauthor{Online-Journal Subcommittee}

\title{%
Stochastic Cutoff Method for Long-Range Interacting Systems
}

\author{%
Munetaka Sasaki and Fumitaka Matsubara}

\inst{%
Department of Applied Physics, Tohoku University, Sendai, 980-8579
}

{\em }\recdate{\today}

\abst{%
A new Monte Carlo method for long-range interacting systems is presented. 
This method involves eliminating interactions stochastically
with the detailed balance condition satisfied. 
When pairwise interactions $V_{ij}$ of an $N$-particle system 
decrease with the distance as $r_{ij}^{-\alpha}$, 
computational time per Monte Carlo step is ${\cal O}(N)$ 
for $\alpha \ge d$ and ${\cal O}(N^{2-\alpha/d})$ for $\alpha < d$, 
where $d$ is the spatial dimension. 
We apply the method to a two-dimensional magnetic dipolar system. 
The method enables us to treat a huge system of $256^2$ spins 
within a reasonable computational time, and reproduces 
a circular order originating from long-range dipolar interactions.}
\kword{%
long-range interaction, Monte Carlo, algorithm, dipolar interactions
}

\maketitle
\newpage
\section{Introduction}
Numerical simulations of long-range interacting systems 
are quite difficult because we have to take a large number of 
interactions into consideration. If one carries out a naive 
Monte Carlo (MC) simulation of an $N$-particle system 
with pairwise interactions, computational time per MC step $t_{\rm MC}$ 
is proportional to $N^2$. Due to this rapid increase in 
computational time, accessible sizes for numerical simulations 
are restricted. One might think that this problem can be resolved 
by truncating interactions beyond a certain cutoff distance. 
However, such truncations often bring significant errors in various 
observables.~\cite{Brooks85,Loncharich89,Schreiber92,Guenot93,Steinbach94,Saito94} Concerning long-range interacting Ising ferromagnetism, 
a cluster algorithm that drastically improves computational efficiency 
without any approximation has been proposed.~\cite{LuijtenBlote95} 
However, this method cannot be used for other systems. 
To overcome this difficulty, some approximate 
methods have been proposed until 
now.~\cite{Appel85,Barnes86,Greengard88,Carrier88,Saito92,Ding92,Sasaki96}
Some of them can treat more than one million particles 
within a reasonable computational time with high accuracy. 
Furthermore, these methods are applicable to 
general long-range interacting systems. 
Nevertheless, these methods include some approximations more or less. 

In this paper, we present a new MC method 
for general long-range interacting systems. 
In contrast to other methods, the present method is {\it exact} 
in the sense that it strictly satisfies 
the detailed balance condition. 
In this method, we stochastically switch long-range interactions $V_{ij}$ 
to either zero or a pseudointeraction $\bar V_{ij}$ 
by use of the stochastic potential switching 
algorithm.~\cite{Mak05,MakSharma07} 
Then the system is mapped on that with only $\bar V_{ij}$. 
The potential switching is performed every several steps. 
Since most of the distant (and weak) interactions are 
eliminated $(V_{ij}\rightarrow 0)$, 
$t_{\rm MC}$ is significantly reduced. 
We refer to the present method as 
the {\it stochastic cutoff (SCO)} method. Of course, if one naively switches 
$V_{ij}$, it costs computational time $t_{\rm switch}$ 
of order $N^2$. We develop an efficient method 
for the potential switching. In lattice systems, it reduces 
$t_{\rm switch}$ to be comparable to $t_{\rm MC}$. 
We apply the SCO method to a two-dimensional magnetic dipolar system. 
%As we will show later, $t_{\rm MC}$ for dipolar systems 
%is of order $N$ in two dimensions and $N\log N$ in three dimensions. 
By comparing our data with the previous ones,~\cite{Sasaki96} 
we confirm that the SCO method gives correct results 
with modest computational time. We discuss the properties
of the SCO method in comparison with other methods for long-range
interacting systems. 

The organization of the paper is as follows. In \S~\ref{sec:method}, 
we describe the SCO method. In \S~\ref{sec:results}, we show the results 
obtained by applying the SCO method to a two-dimensional 
magnetic dipolar system. In \S~\ref{sec:comparison}, 
we compare the SCO method with other methods. 
Section~\ref{sec:conclusion} is devoted to conclusions. 

\section{Stochastic Cutoff (SCO) Method}
\label{sec:method} 
Before explaining the SCO method, we briefly survey 
the stochastic potential switching algorithm.~\cite{Mak05,MakSharma07} 
We hereafter consider a lattice system with pairwise 
long-range interactions described by the Hamiltonian
\begin{equation}
{\cal H}=\sum_{i<j}V_{ij}(\vectr{S}_i,\vectr{S}_j),
\label{eqn:GH}
\end{equation} 
where $\vectr{S}_i$ is a variable associated with the $i$-th element 
of the system. In this algorithm, $V_{ij}$ is stochastically switched 
to either $\tilde{V}_{ij}$ or $\bar{V}_{ij}$ 
with a probability of $P_{ij}$ or $1-P_{ij}$, respectively. 
The probability $P_{ij}$ is 
\begin{equation}
P_{ij}(\vectr{S}_i,\vectr{S}_j)=
\exp[\beta (\Delta V_{ij}(\vectr{S}_i,\vectr{S}_j)-\Delta V_{ij}^*)],
\label{eqn:Sprob}
\end{equation}
where $\beta$ is the inverse temperature, 
$\Delta V_{ij}(\vectr{S}_i,\vectr{S}_j)\equiv V_{ij}(\vectr{S}_i,\vectr{S}_j)-
\tilde V_{ij} (\vectr{S}_i,\vectr{S}_j)$, 
and $\Delta V_{ij}^*$ is a constant equal to (or greater than) 
the maximum value of $\Delta V_{ij}(\vectr{S}_i,\vectr{S}_j)$ over all 
$\vectr{S}_i$ and $\vectr{S}_j$. We can choose the potential 
$\tilde{V}_{ij}$ {\it arbitrarily}. 
On the other hand, using $P_{ij}(\vectr{S}_i,\vectr{S}_j)$, 
the potential $\bar{V}_{ij}$ is given as 
\begin{equation}
\bar{V}_{ij}(\vectr{S}_i,\vectr{S}_j)=
V_{ij}(\vectr{S}_i,\vectr{S}_j)
-\beta^{-1}\log[1-P_{ij}(\vectr{S}_i,\vectr{S}_j)].
\label{eqn:Ppseudo}
\end{equation}
The arbitrariness of $\tilde{V}_{ij}$ can be utilized to 
reduce either the complexity of potential or cost to calculate it. 
With this potential switching process, the algorithm proceeds as follows:
\begin{itemize}
\item[(A)] Potentials $V_{ij}$ are switched 
to either $\tilde{V}_{ij}$ or $\bar{V}_{ij}$ with the probability of 
$P_{ij}$ or $1-P_{ij}$, respectively.
\item[(B)] A standard MC simulation is performed with the 
switched Hamiltonian ${\cal H}'$ expressed as 
\begin{equation}
{\cal H}'=\sum\nolimits_{ij}'\tilde{V}_{ij}(\vectr{S}_i,\vectr{S}_j)+
\sum\nolimits_{ij}''\bar{V}_{ij}(\vectr{S}_i,\vectr{S}_j),
\label{eqn:Spotential}
\end{equation}
where $\sum'$ runs over all the potentials switched to $\tilde{V}$ 
and $\sum''$ runs over those switched to $\bar{V}$. 
The potential is fixed during the simulation. 
\item[(C)] Return to (A).
\end{itemize}
It is shown that this MC procedure strictly satisfies 
the detailed balance condition with respect to the original Hamiltonian 
of eq.~(\ref{eqn:GH}). 

We give two remarks. 
Firstly, we can choose the period of the simulation in step (B) 
{\it arbitrarily} because the detailed balance condition is satisfied 
regardless of the period. Secondly, we do not need to switch 
all the potentials. For example, when the original Hamiltonian is
${\cal H}=\sum\nolimits_{ij} V_{ij}+\sum\nolimits_{ij} U_{ij},$
we can switch only $\{ V_{ij} \}$ with $\{ U_{ij}\}$ unswitched. 
Such partial switching is realized by setting $\tilde U_{ij} =U_{ij}$.

We now describe the SCO method. The basic idea is quite simple. 
We just set $\tilde{V}_{ij}=0$ to reduce computational time. 
In the following, we see that time $t_{\rm MC}$ in step (B) 
is significantly reduced since most potentials associated 
with distant interactions are switched to $\tilde{V}_{ij}=0$ in step (A). 
%Furthermore, we can reduce time required in step (A) dramatically 
%by using an efficient method. 
%Now let us first evaluate time needed in step (B). 
Now let us assume that interactions $V_{ij}$ decrease 
as $D r_{ij}^{-\alpha}$ ($\alpha > 0$), where $D$ is a constant 
that represents the strength of interactions and $r_{ij}$ is the distance 
between sites $i$ and $j$. Systems with dipolar interactions 
correspond to the case $\alpha=3$. When we update an element 
from $\vectr{S}_i^{({\rm old})}$ to $\vectr{S}_i^{({\rm new})}$ 
in a MC simulation, we need to calculate the energy difference 
\begin{equation}
\Delta {\cal H}'_i\equiv\sum\nolimits_{k}' \left[{\bar V}_{ik}
\left(\vectr{S}_i^{({\rm new})},\vectr{S}_k\right)
-{\bar V}_{ik}\left(\vectr{S}_i^{({\rm old})},\vectr{S}_k\right)\right],
\end{equation}
where the sum is taken over all the sites $k$ for which 
$V_{ik}$ is switched to ${\bar V}_{ik}$. 
This means that the time required to update $\vectr{S}_i$ is 
proportional to the number of terms, i.e., ${\cal N}_i\equiv \sum_k'$.  
When $r_{ik}$ is large, the probability that $V_{ik}$ is switched 
to ${\bar V}_{ik}$ is approximated as
\begin{equation}
1-P_{ik} \sim D\beta r_{ik}^{-\alpha}
\quad\left(r_{ik}\gg\left( D\beta \right)^{1/\alpha}\right),
\label{eqn:Sapprox}
\end{equation}
where we have assumed that both $\Delta V_{ik}$ and
$\Delta V_{ik}^*$ are of order $Dr_{ik}^{-\alpha}$.  
Therefore, we can roughly evaluate ${\cal N}_i$ as
\begin{eqnarray}
{\cal N}_i&\sim& \int_1^{L} {\rm d}r r^{d-1} D\beta r^{-\alpha}\nonumber\\
&\sim& \left\{
\begin{array}{cc}
D\beta & (\alpha> d),\\
D\beta \log(L) &  (\alpha=d),\\
D\beta L^{d-\alpha} & (\alpha<d), 
\end{array}
\right.
\label{eqn:Ni_esti}
\end{eqnarray}
where $d$ is the spatial dimension of the system, $L$ is the linear 
size of the lattice, and the lattice constant is assumed to be one. 
Therefore, computational time per one MC step $t_{\rm MC}$ 
is estimated as
\begin{equation}
t_{\rm MC} \propto N{\cal N}_i
\sim \left\{
\begin{array}{cc}
D\beta N & (\alpha>d),\\
D\beta N \log(N) &  (\alpha=d),\\
D\beta N^{2-\alpha/d} & (\alpha<d),\\
\end{array}
\right.
\label{eqn:tMC}
\end{equation}
where $N(=L^d)$ is the total number of elements and 
the difference between $\log(L)$ and $\log(N)$ is ignored. 
Obviously, $t_{\rm MC}$ in the present method 
is much smaller than that in a naive MC method 
where $t_{\rm MC}\sim N^2$. 

\begin{figure}
\begin{center}
\includegraphics[width=14cm]{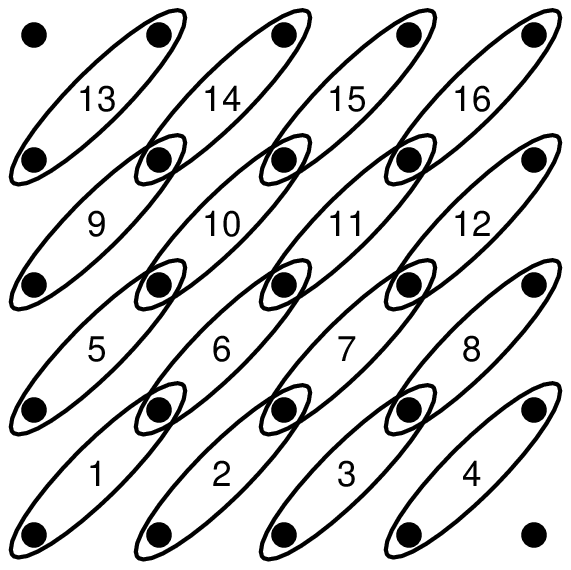}
\end{center}
\caption{}
\label{fig:pairs}
\end{figure}

We next consider the potential switching process. 
If one switches $V_{ij}$ one by one, the potential switching time 
$t_{\rm switch}$ is of order $N^2$. 
To reduce $t_{\rm switch}$, 
we develop the following method. 
%In lattice systems, it reduces $t_{\rm switch}$ to be comparable to 
%$t_{\rm MC}$ given by eq.~(\ref{eqn:tMC}). 
We first introduce a set of pairs 
$\{ {\cal C}_{\rm pair}(\vectr{r}) \}$ for which either $\vectr{r}_{ij}$ 
or $\vectr{r}_{ji}$ is $\vectr{r}$, where $\vectr{r}_{ij}$ is the vector 
spanning from sites $i$ to $j$. 
Figure~\ref{fig:pairs} shows an example of such sets. 
Pairs are labeled sequentially. 
The point is that, in most cases, the probabilities of switching
to $\bar V$ 
for $\{ {\cal C}_{\rm pair}(\vectr{r}) \}$ have some upper limit 
$p_{\rm max}(\vectr{r})$. 
As we will show later, such maximum probability 
indeed exists in dipolar systems. 
%For example, in the case of dipolar 
%interactions (the second term in the right-hand of 
%eq.~(\ref{eqn:Hamiltonian1})), $V_{ij}(=\Delta V_{ij})$ has 
%its minimum (maximum) value when $\vectr{S}_i$ and $\vectr{S}_j$ are 
%in parallel (anti-parallel) along $\vectr{r}_{ij}$. The minimum value 
%is $-2D/r_{ij}^3$ and the maximum value is $+2D/r_{ij}^3$. Therefore, 
%when we set $\Delta V^*_{ij}$ to be $2D/r_{ij}^3$, 
%we obtain $p_{\rm max}(\vectr{r}) = 1-\exp[-4D\beta /r^3]$ since 
%\begin{eqnarray}
%1-P_{ij}(\vectr{S}_i,\vectr{S}_j)&=&
%1-\exp[\beta \{V_{ij}(\vectr{S}_i,\vectr{S}_j)-2D/r_{ij}^3\}]\nonumber\\
%&\le&1-\exp[-4D\beta /r_{ij}^3].
%\end{eqnarray}
Using $p_{\rm max}(\vectr{r})$, we can switch the potentials in 
$\{ {\cal C}_{\rm pair}(\vectr{r}) \}$ in the following way: 
\begin{itemize}
\item[a)] Using $p_{\rm max}(\vectr{r})$, we choose {\it candidates} 
that are switched to ${\bar V}_{ij}$. 
The potentials that are not chosen as a candidate are switched to $0$. 
\item[b)] Switch each of the candidates to ${\bar V}_{ij}$ with the probability
$[1-P_{ij}(\vectr{S}_i,\vectr{S}_j)]/p_{\rm max}(\vectr{r})$. Otherwise, 
$V_{ij}$ is switched to $0$. 
\end{itemize}
Of course, if candidates are chosen one by one, 
it is very timeconsuming. We therefore choose them 
in the following way. We hereafter denote the potential of 
the $k$-th pair in $\{ {\cal C}_{\rm pair}(\vectr{r}) \}$ 
by $V^{(k)}$. Since the probability of being chosen as a candidate 
is the same in $\{ {\cal C}_{\rm pair}(\vectr{r}) \}$, 
the probability that $V^{(n)}$ is chosen as a candidate 
after $n-1$ successive failures is given 
by a geometric distribution $g(p_{\rm max}(\vectr{r}),n)$, where
\begin{equation}
g(p,n)=(1-p)^{n-1}p\quad(n\ge 1). 
\end{equation}
An integer random variate $n$ that obeys $g(p,n)$ can be easily generated as
\begin{equation}
n=\left\lceil \frac{\log(r)}{\log(1-p)}\right\rceil, 
\label{eqn:GDgne}
\end{equation}
where $\lceil x \rceil$ is the smallest integer that is greater than or 
equal to $x$, and $r$ is a continuous random variate with an uniform 
distribution of range $0<r\le 1$. We can pick up 
only candidates by means of $g(p_{\rm max}(\vectr{r}),n)$. For example, 
the generation of two random variates $n_1$ and $n_2$ means 
that there are only two candidates $V^{(n_1)}$ and $V^{(n_1+n_2)}$ 
among $n_1+n_2$ potentials. Using this idea, 
we have implemented the potential switching in 
$\{ {\cal C}_{\rm pair}(\vectr{r}) \}$ as follows:
\begin{itemize}
\item[1)] Set $n_{\rm s}$ to be $0$, where $n_s$ is the number of 
potentials that have already been switched. 
\item[2)] Generate an integer $n$ from the distribution 
$g(p_{\rm max}(\vectr{r}),n)$ using eq.~(\ref{eqn:GDgne}). 
If $n=1$, go to step 4). Otherwise, go to the next step. 
\item[3)] Switch the $n-1$ potentials 
($V^{(n_{\rm s}+1)}$, $V^{(n_{\rm s}+2)}$, $\cdots$, $V^{(n_{\rm s}+n-1)}$) 
to $0$. 
%\item[4)] Finish the potential switching procedure 
%if $n_{\rm s}+n$ is greater than the number of elements of 
%$\{ {\cal C}_{\rm pair}(\vectr{r}) \}$. 
\item[4)] Switch $V^{(n_{\rm s}+n)}$ to ${\bar V}^{(n_{\rm s}+n)}$ 
with the probability 
$[1-P^{(n_{\rm s}+n)}(\vectr{S}_i,\vectr{S}_j)]/p_{\rm max}(\vectr{r})$. 
Otherwise, switch it to $0$. 
\item[5)] Finish the potential switching procedure 
if $n_{\rm s}+n$ is greater than (or equal to) the number of elements of 
$\{ {\cal C}_{\rm pair}(\vectr{r}) \}$. Otherwise, 
replace $n_{\rm s}$ with $n_{\rm s}+n$ and return to step 2). 
\end{itemize}
Switching of all the potentials is completed by carrying out 
this procedure for all $\{ {\cal C}_{\rm pair}(\vectr{r}) \}$. 

Now let us evaluate $t_{\rm switch}$. The potential switching process 
clearly requires time proportional to the number of potentials chosen 
as a candidate.  To estimate it, we focus on the $N-1$ potentials 
associated with a certain site $i$ and estimate the number of candidates 
${\cal N}_i^{(c)}$ among them. The total number of candidates is 
$N {\cal N}_i^{(c)}$. Since both $p_{\rm max}(\vectr{r})$ and 
$1-P_{ik}$ are of order $D\beta r^{-\alpha}$ 
when $r\gg (D \beta)^{1/\alpha}$, the order estimation of 
${\cal N}_i^{(c)}$ and that of ${\cal N}_i$ given by 
eq.~(\ref{eqn:Ni_esti}) are the same. We therefore obtain
\begin{equation}
t_{\rm switch}
\sim \left\{
\begin{array}{cc}
D\beta N & (\alpha>d), \\
D\beta N \log(N) &  (\alpha=d),\\
D\beta N^{2-\alpha/d} & (\alpha<d).\\
\end{array}
\right.
\label{eqn:tswitch}
\end{equation}
From eqs.~(\ref{eqn:tMC}) and (\ref{eqn:tswitch}), we find that 
$t_{\rm switch}$ is indeed comparable to $t_{\rm MC}$. 
%Because it costs time of ${\cal O}(1)$ even if no candidate is chosen 
%from $\{ {\cal C}_{\rm pair}(\vectr{r}) \}$, 
%the switching process requires some additional time. However, since 
%the number of sets is of order $N$ for lattice systems, 
%the order estimation of eq.~(\ref{eqn:tswitch}) is not changed 
%by taking this additional time into account. 

\section{Results}
\label{sec:results}
Now we apply the SCO method to a two-dimensional magnetic dipolar system 
on an $L\times L$ square lattice with open boundaries. 
The Hamiltonian of the system is described as
\begin{eqnarray}
{\cal H}&&\hspace{-7mm}=-J\sum\nolimits_{\langle ij\rangle}\vectr{S}_i \cdot \vectr{S}_j\nonumber\\
&&\hspace*{-5mm}
+D\sum\nolimits_{i<j}\left[ \frac{\vectr{S}_i\cdot \vectr{S}_j}{r_{ij}^3}
-3\frac{(\vectr{S}_i\cdot\vectr{r}_{ij})(\vectr{S}_j\cdot\vectr{r}_{ij})}
{r_{ij}^5}\right],\nonumber\\
\label{eqn:Hamiltonian1}
\end{eqnarray}
where $\vectr{S}_i$ is a classical Heisenberg spin of $|\vectr{S}_i|=1$, 
$\langle ij \rangle$ runs over all the nearest-neighboring pairs, 
$\vectr{r}_{ij}$ is the vector spanned from sites $i$ 
to $j$ in the unit of the lattice constant $a$, and $r_{ij}=|\vectr{r}_{ij}|$. 
The first term describes short-range ferromagnetic exchange interactions and  
the second term describes long-range dipolar interactions, 
where $J(>0)$ is an exchange constant and $D=(g\mu_{\rm B}S)/a^3$.
Hereafter, we regard $D$ as a parameter and consider the case that 
$D/J=0.1$. We choose this model as a benchmark of the SCO method 
because the properties of the model have been investigated extensively 
in previous work.~\cite{Sasaki96} 
In particular, it is established that the model 
undergoes a phase transition from the paramagnetic state 
to a circularly ordered state at $T_{\rm c}\approx 0.88J$ as a consequence 
of the cooperation of exchange and dipolar interactions. 

Before showing the results, we explain the details of our simulation. 
We applied the SCO method only for dipolar interactions. 
The potential difference $\Delta V_{ij}$ is given as
\begin{equation}
\Delta V_{ij}=V_{ij}=D\left[\frac{\vectr{S}_i\cdot \vectr{S}_j}{r_{ij}^3}
-3\frac{(\vectr{S}_i\cdot\vectr{r}_{ij})(\vectr{S}_j\cdot\vectr{r}_{ij})}
{r_{ij}^5}\right].
\end{equation}
It has the minimum value $-2D/r_{ij}^3$ and the maximum value $+2D/r_{ij}^3$
when $\vectr{S}_i$ and $\vectr{S}_j$ are 
parallel along $\vectr{r}_{ij}$ and antiparallel, respectively. 
Therefore, we obtain
\begin{eqnarray}
&&\hspace*{-5mm}p_{\rm max}(\vectr{r}) \nonumber\\
&&\hspace*{-5mm}=\max\nolimits_{\vectr{S}_i,\vectr{S}_j}\Bigl\{1-\exp[\beta(\Delta V_{ij}
(\vectr{S}_i,\vectr{S}_j)-\Delta V_{ij}^*)] \Bigr\}\nonumber\\
&&\hspace*{-5mm}=1-\exp[-4D\beta/r^3],
\end{eqnarray}
where we have set $\Delta V_{ij}^*=2D/r^3$. 
The system was gradually cooled from an initial temperature $T=1.65J$ 
to $0.05J$ in steps of $\Delta T=0.05J$. The initial temperature was 
set to be well above the critical temperature. 
The system was kept at each temperature for $100,000$ MC steps, 
and potentials were switched for every $10$ MC steps.  
The first $50,000$ MC steps are for equilibration and the following 
$50,000$ MC steps are for measurement. Therefore, the total MC steps 
for one run is $3,300,000$. The computational time per run for $L=256$, 
i.e, the maximum size we examined, was less than three days 
when using a personal computer with a Core2Duo 2.40 GHz processor. 
We conducted simulations for $10$ different runs 
with different initial conditions and random sequences. 
The energy was calculated for every $10$ MC steps with ${\cal O}(N\log N)$
computational time by utilizing the discrete Fourier convolution theorem and 
the fast Fourier transformation algorithm. 
For a detailed description of how the discrete Fourier convolution theorem 
is used for a system with open boundary conditions, we refer the reader to 
ref.~\citen{Sasaki96}.

%The Hamiltonian (\ref{eqn:Hamiltonian1}) is rewritten as 
%\begin{equation}
%{\cal H}=-\frac{1}{2} \sum_{i\ne j}\left\{\sum_{\mu,\nu}
%f^{(\mu\nu)}(\vectr{r}_{ij})S_i^{\mu}S_j^{\nu}\right\},
%\end{equation}
%where $\mu,\nu=x,y,z$ and 
%\begin{equation}
%f^{(\mu,\nu)}(\vectr{r})=J\sum_{\vectr{a}\in n.n.}\delta_{\vectr{a},\vectr{r}}
%-D\left\{\frac{\delta_{\mu\nu}}{r^3}-3\frac{r^{\mu}r^{\nu}}{r^5} \right\}. 
%\end{equation}
%Therefore, from the discrete Fourier convolution theorem, we obtain
%\begin{equation}
%{\cal H}=\sqrt{N}\sum_{\vectr{k}}\sum_{\mu,\nu}{\tilde f}^{(\mu,\nu)}
%(\vectr{k}) {\tilde S}^{\mu}(\vectr{k}){\tilde S}^{\nu}(\vectr{-k}),
%\end{equation}
%where ${\tilde f}^{(\mu,\nu)}$ and ${\tilde S}^{\mu}$ are the Fourier 
%coefficient of $f^{(\mu,\nu)}$ and that of $S^{\mu}$, respectively. 
%By combining this equation with the fast Fourier transformation algorithm, 
%we can evaluate ${\cal H}$ with computational time of order $N\log N$. 

\begin{figure}
\begin{center}
\includegraphics[angle=270,width=16cm]{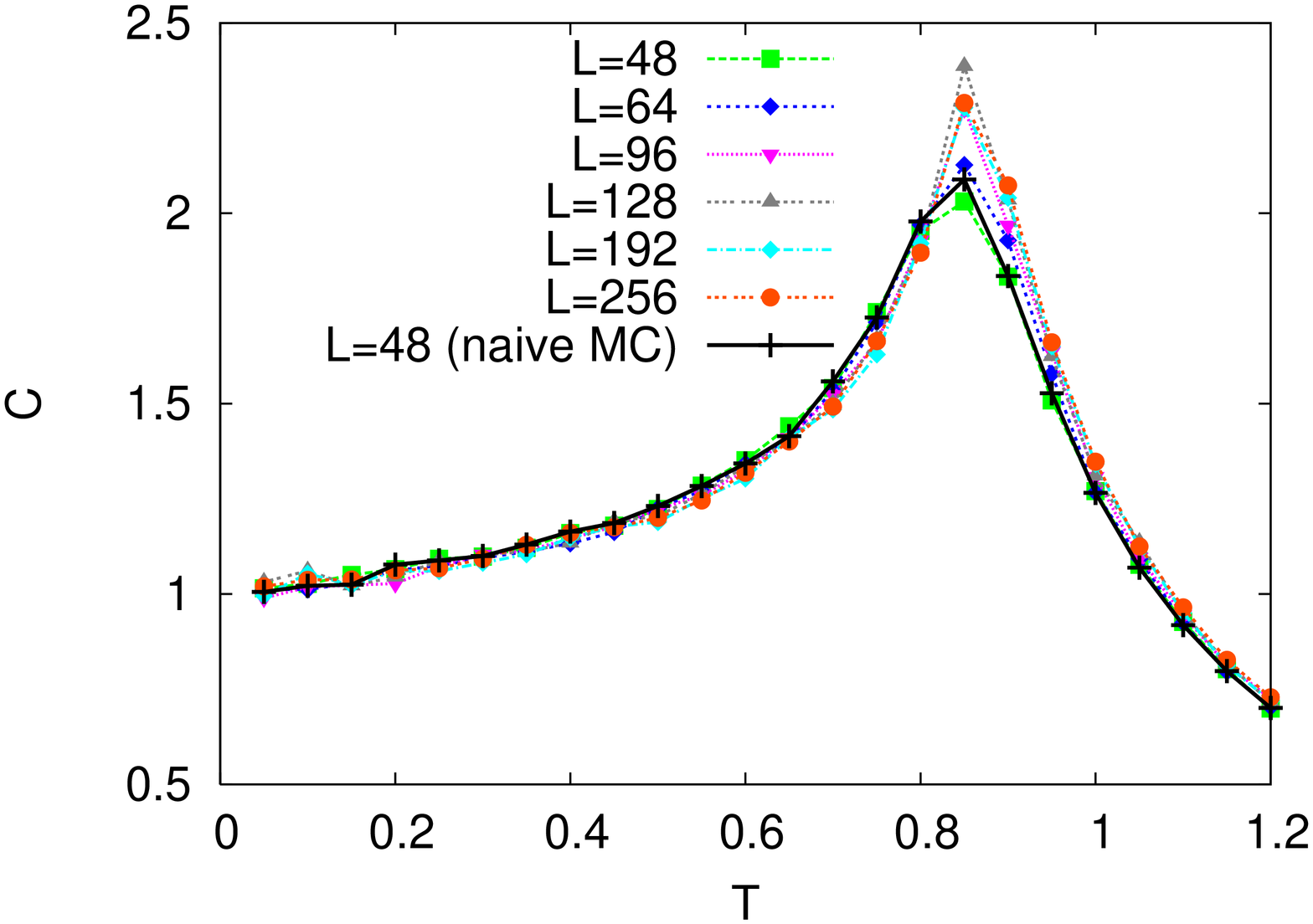}
\end{center}
\caption{}
\label{fig:capacity}
\end{figure}

\begin{figure}
\begin{center}
\includegraphics[width=16cm]{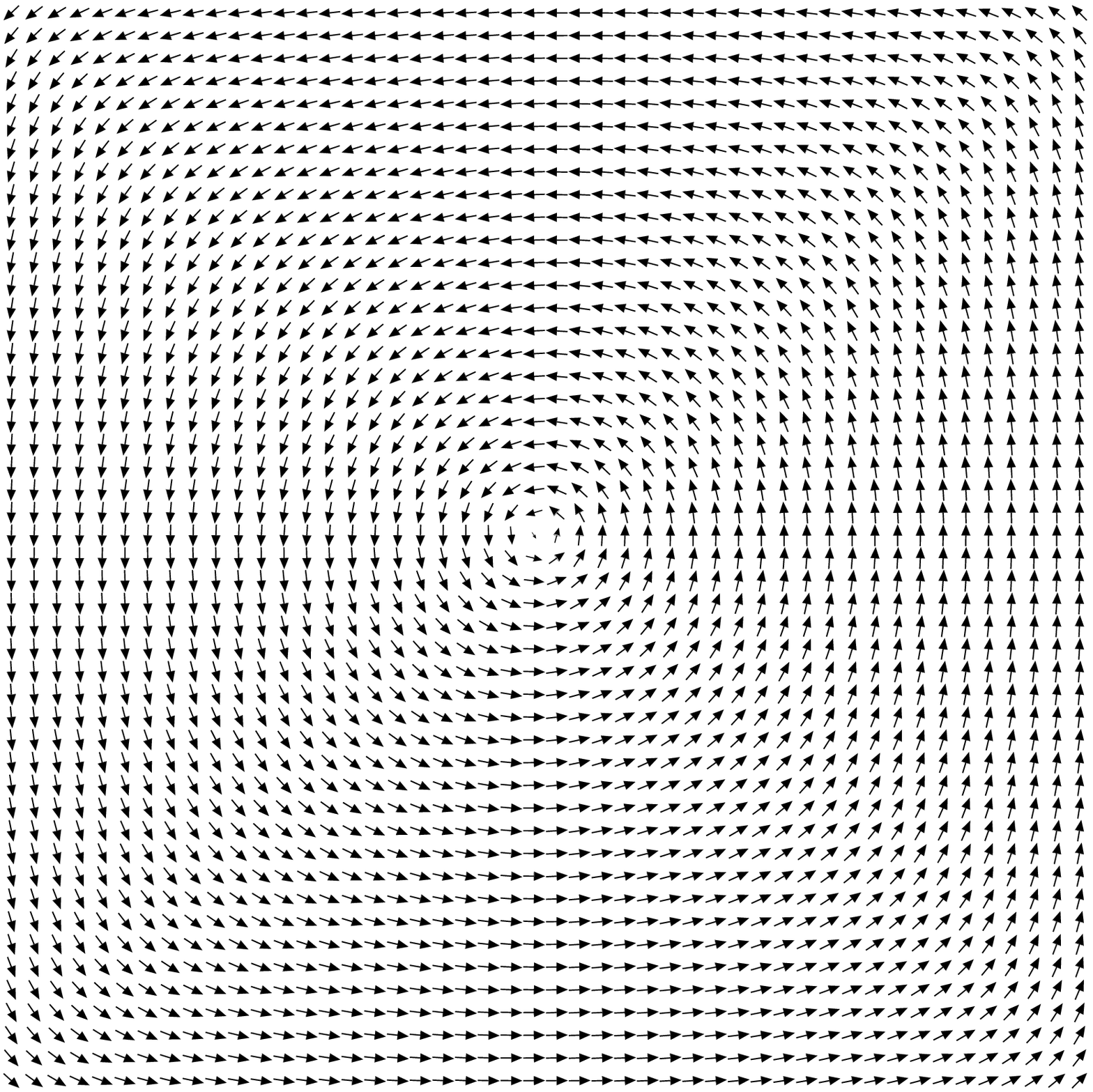}
\end{center}
\caption{}
\label{fig:snapshot1}
\end{figure}

\begin{figure}
\begin{center}
\includegraphics[angle=270,width=16cm]{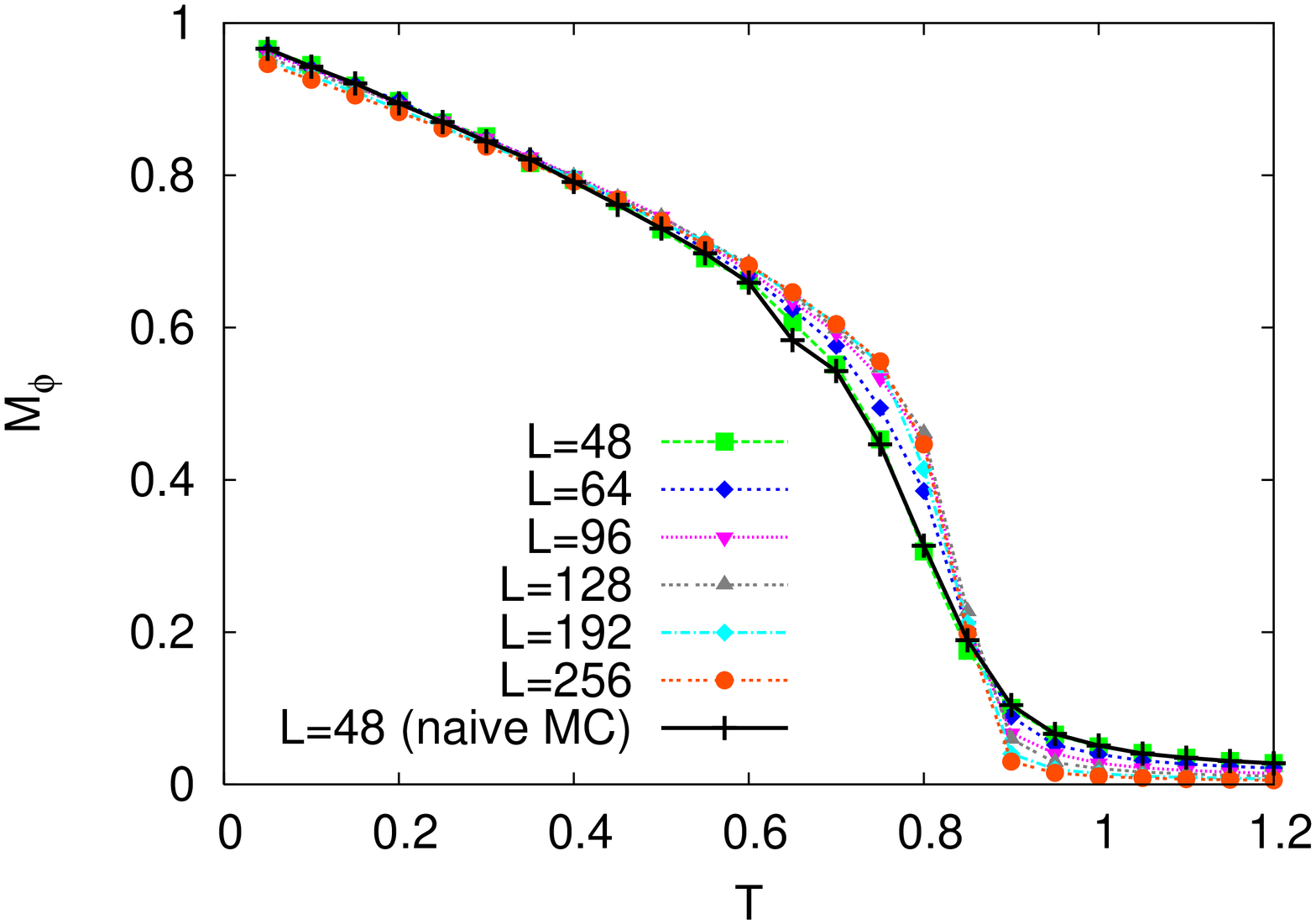}
\end{center}
\caption{}
\label{fig:Mc}
\end{figure}

Now let us see the results of our simulations. Figure~\ref{fig:capacity} 
shows the temperature dependences of the specific heat for different sizes. 
The peaks are located around $T_{\rm c}\approx 0.88J$. 
In Fig.~\ref{fig:snapshot1}, we show the spin structure observed 
at $T=0.05J$. We clearly see a circular order 
that comes from long-range dipolar interactions. 
For a quantitative measurement of the circular order, we observed 
the absolute value of the circular component defined by
\begin{equation}
M_{\phi} \equiv \left\langle \left| \left[ \frac{1}{N} \sum_i 
\vectr{S}_i \times \frac{\vectr{r}_i-\vectr{r}_{\rm c}}
{|\vectr{r}_i-\vectr{r}_{\rm c}|}
\right]\right|_z\right\rangle,
\end{equation}
where $\langle \cdots\rangle$ denotes the thermal average 
and $\vectr{r}_{\rm c}$ is a vector describing the center 
of the lattice. 
Figure~\ref{fig:Mc} shows the result. We see that 
the circular order rapidly grows around the critical temperature. 
We also performed a naive MC simulation for $L=48$ to confirm that 
the SCO method reproduces correct results. 
The crosses in Figs.~\ref{fig:capacity}~and~\ref{fig:Mc} show the results. 

\begin{figure}
\begin{center}
\includegraphics[angle=270,width=16cm]{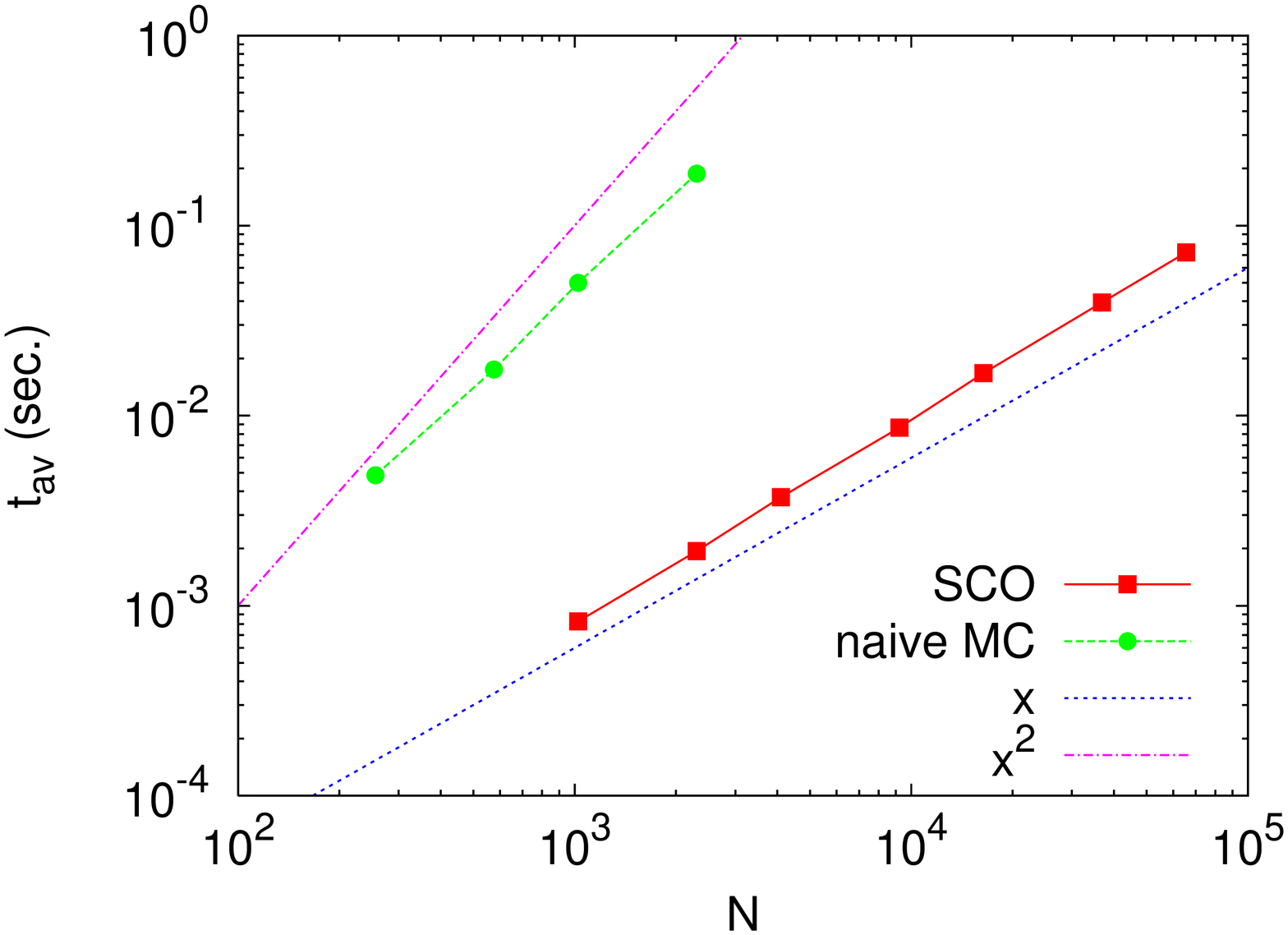}
\end{center}
\caption{}
\label{fig:time}
\end{figure}

Now let us examine the efficiency of the SCO method. Figure~\ref{fig:time} 
shows the size dependences of the average computational time 
per MC step $t_{\rm av}$ for both the SCO method and the naive MC method. 
The average time $t_{\rm av}$ of the SCO method is given as
\begin{equation}
t_{\rm av}=t_{\rm MC}+\frac{1}{10}t_{\rm switch}+\frac{1}{10}t_{\rm energy},
\end{equation}
where $t_{\rm energy}$ is the computational time per one energy measurement. 
Recall that potential switching and energy measurement are performed for every 
$10$ MC steps. We see that $t_{\rm av}\propto N$ in the SCO method, which 
is strongly in contrast with $t_{\rm av}\propto N^2$ in the naive MC method.

\begin{figure}
\begin{center}
\includegraphics[angle=270,width=16cm]{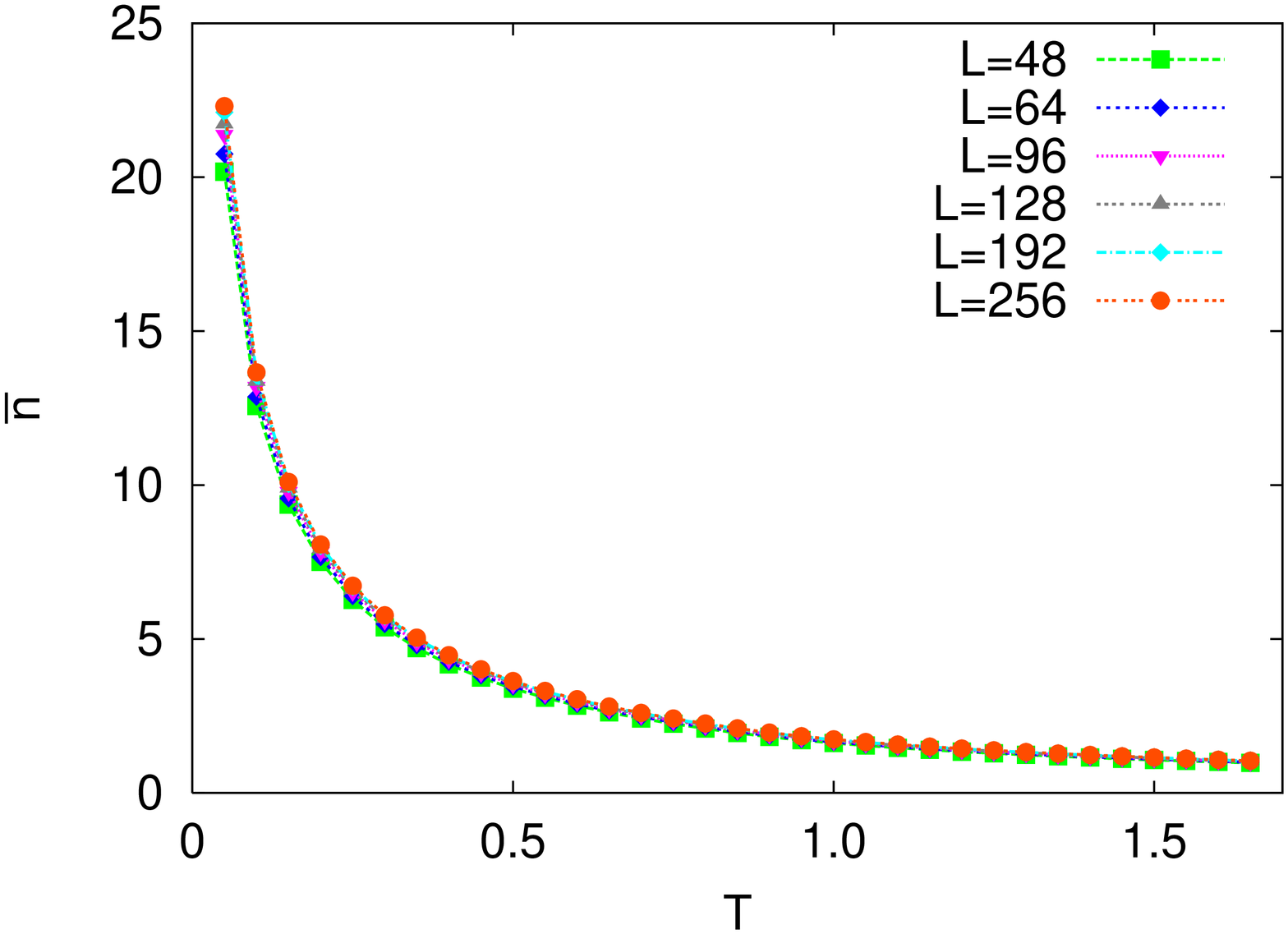}
\end{center}
\caption{}
\label{fig:number}
\end{figure}

This huge reduction of the computational time in the SCO method 
comes from the reduction of interactions. We 
observed the average number $\bar{n}$ of potentials per site 
that survive as $\bar V_{ij}$. 
In Fig.~\ref{fig:number}, we show the temperature dependences of $\bar n$ 
for different sizes. It is impressive that $\bar n$ at each temperature 
converges to a certain value as the size increases. 
%Note that $\alpha=3$ and $d=2$ in the present case. 
Although $\bar n$ increases with decreasing temperature, 
$\bar n\sim22.5$ even at the lowest temperature. 
%This means that the computational time of the present method is 
%close to that of a naive cutoff method with the cutoff distance of 
%$\sqrt{22.5/\pi}\approx 2.7$. However, as demonstrated 
%in Fig.~\ref{fig:snapshot2}, cutoff method with such a short cutoff 
%distance does not reproduce the circular order. 

\begin{figure}
\begin{center}
\includegraphics[angle=270,width=16cm]{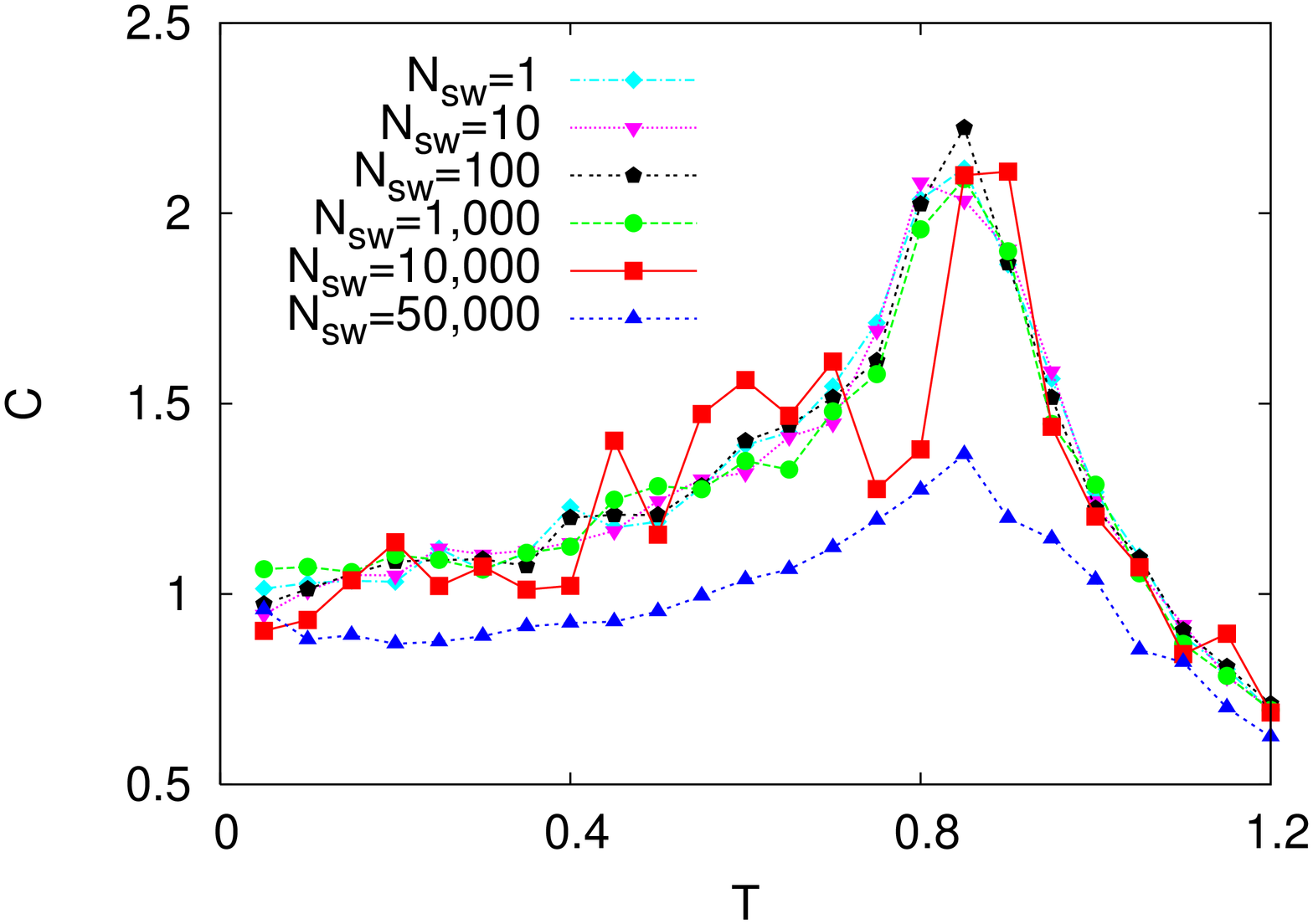}
\end{center}
\caption{}
\label{fig:KDWdependence}
\end{figure}

\begin{figure}
\begin{center}
\includegraphics[angle=270,width=16cm]{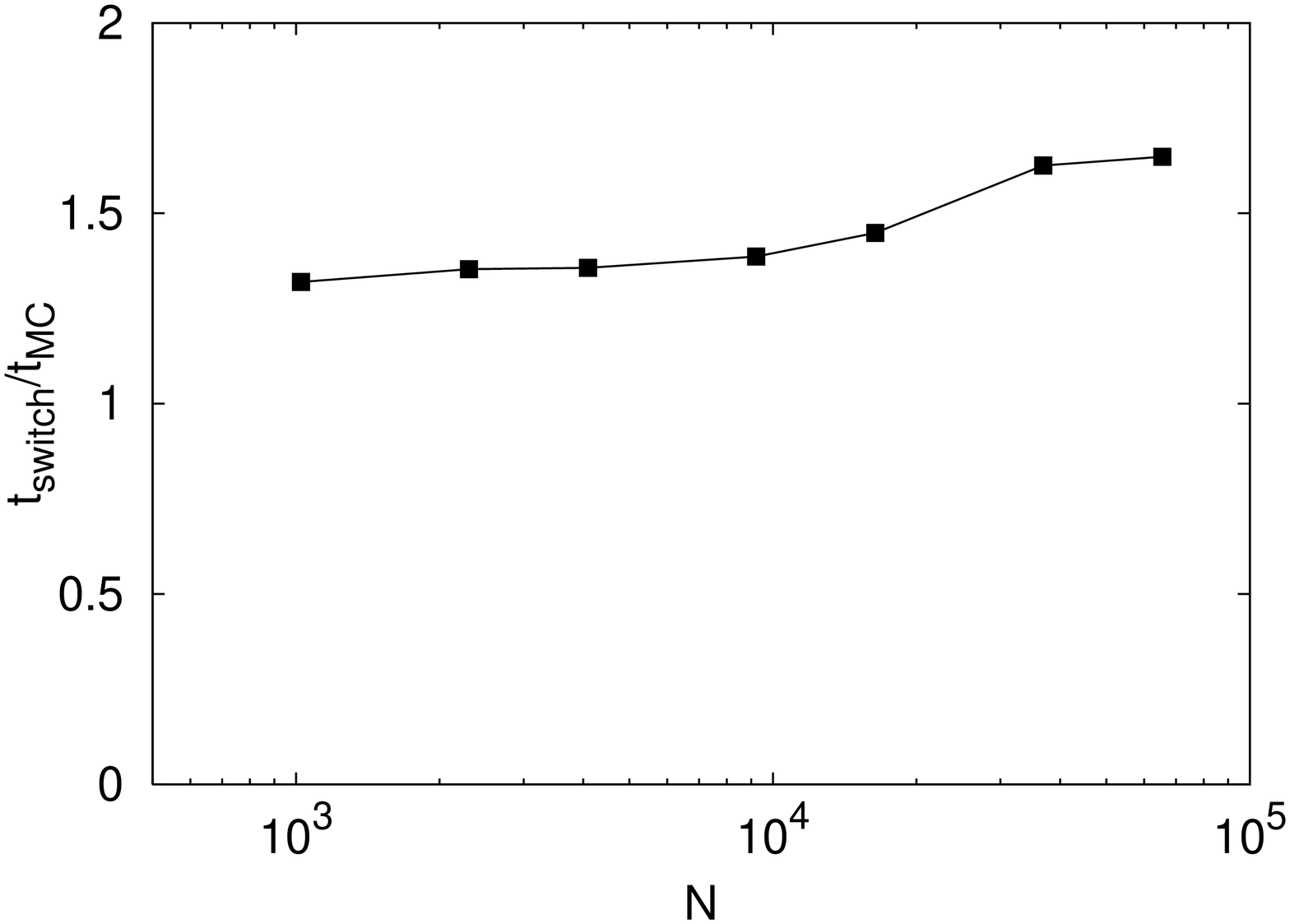}
\end{center}
\caption{}
\label{fig:ratio}
\end{figure}

In Fig.~\ref{fig:KDWdependence}, we show the data of the specific heat 
measured in several simulations with different potential switching 
periods $N_{\rm sw}$. 
Since the system is kept at each temperature for $100,000$ MC steps, 
the number of potential switchings at each temperature is 
$100,000/N_{\rm sw}$. We see that reliable results are obtained when 
$N_{\rm sw}$ is $1,000$ or less. Figure~\ref{fig:ratio} shows 
the size dependence of the ratio $t_{\rm switch}/t_{\rm MC}$. 
The ratio slightly depends on the size, and it is $1.65$ at most. 
This means that we can even switch potentials 
at every MC step with a reasonable cost. 

\begin{figure}
\begin{center}
\includegraphics[angle=270,width=16cm]{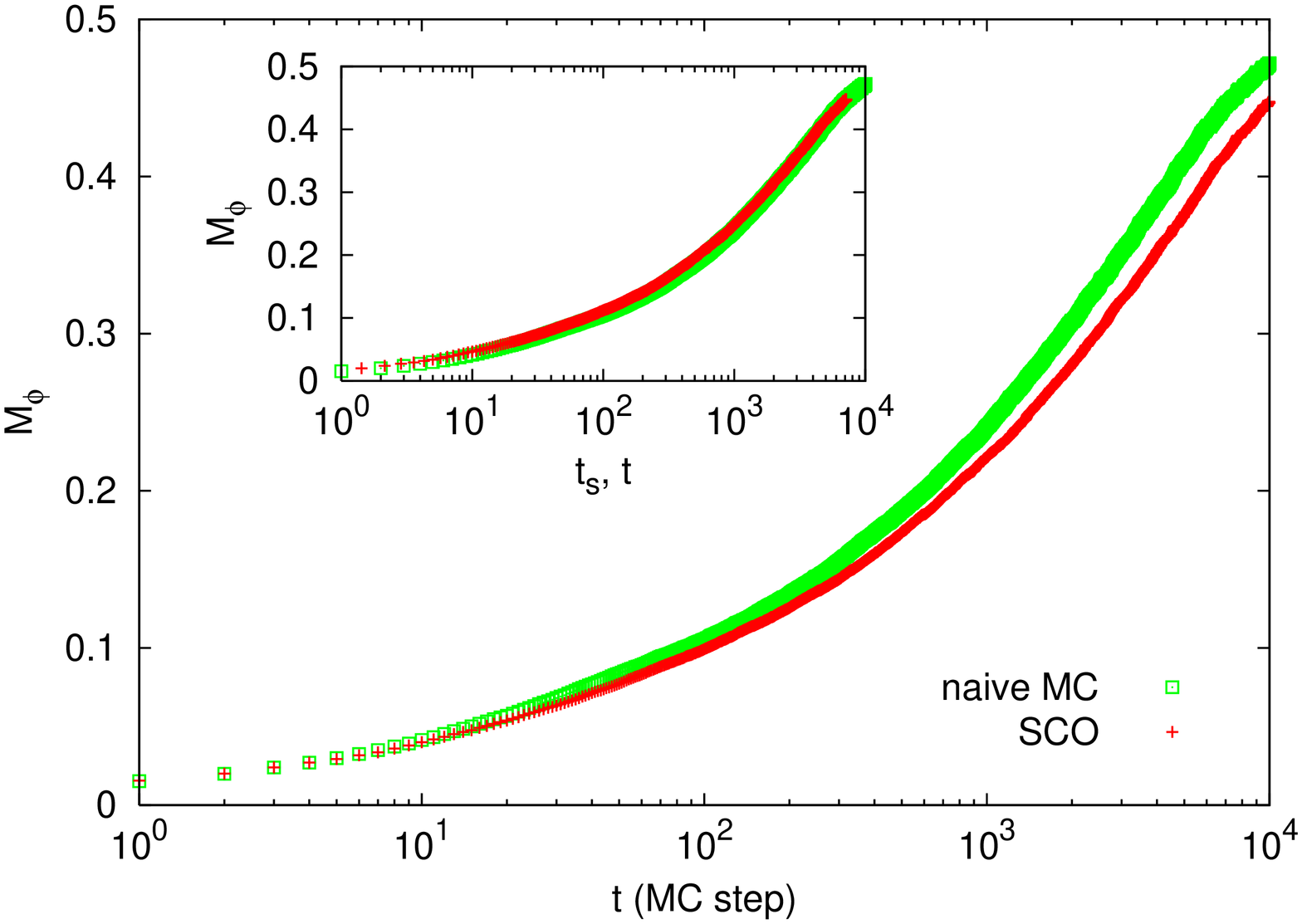}
\end{center}
\caption{}
\label{fig:relax}
\end{figure}

Lastly, we compare relaxation speeds between the SCO method 
and the naive MC method. Figure~\ref{fig:relax} shows the time evolution of 
$M_{\phi}$ for $L=48$ when the system is kept at $T=0.4J$. 
As shown in the inset, the relaxation speed of the SCO method is about 
$1.4$ times slower than that of the naive MC method. We have also performed 
a similar measurement at $T=0.7J$, and found that the ratio is 
about $1.2$. It is clear from Fig.~\ref{fig:time} that
the relaxation speed of the SCO method is much higher than that of 
the naive MC method if they are compared in terms of the computational time.

%\textcolor{blue}{
%Lastly, Fig.~\ref{fig:snapshot2} shows the spin structure at $T=0.05J$ 
%obtained by a naive cutoff method. In order to make the 
%computational time comparable with that of the SCO method, 
%we set the cutoff distance to be $\sqrt{22.5/\pi}\approx 2.7$. 
%Recall that $\bar n\approx 22.5$ at the lowest temperature. 
%The figure clearly shows that the circular order is not reproduced. 

\section{Comparison of the SCO Method with Other Methods}
\label{sec:comparison}
In this section, we discuss the properties of the SCO method 
in comparison with those of other methods for long-range interacting systems. 
As we have emphasized, the primary merit of the SCO method is that 
it involves no approximation. The performance of this method 
strongly depends on the conditions of simulations in terms of parameters 
such as the temperature $T$, the spatial dimension $d$, 
the decay exponent of potentials $\alpha$, 
and the strength of potentials $D$, as illustrated 
by eqs.~(\ref{eqn:tMC}) and (\ref{eqn:tswitch}). 
Since $t_{\rm MC}$ and $t_{\rm switch}$ 
are proportional to $D\beta$, 
the SCO method is particularly efficient for systems with {\it strong} 
short-range interactions and {\it weak} long-range interactions. 
The reason is as follows: 
Since the system is dominated by short-range interactions, 
we expect $k_{\rm B}T_{\rm c}\sim D_{\rm SR}$, where 
$D_{\rm SR}$ is the strength of short-range interactions. 
This means that $t_{\rm MC}$ and $t_{\rm switch}$ are 
very small around the critical temperature because
\begin{equation}
D\beta_{\rm c}\sim D/D_{\rm SR} \ll 1.
\end{equation}
In this sense, the SCO method is really suitable for magnetic dipolar systems, 
as we have demonstrated. In magnetic 
dipolar systems, the ratio $J/D$ is usually on the order of hundreds 
or even thousands. On the other hand, when $\alpha < d$, 
the computational time in the SCO method is 
${\cal O}(N^{2-\alpha/d})$. The SCO method is less efficient in such a case
because the computational time is of order $N$ or $N\log N$
in most of the other methods.

Lastly, we emphasize that the SCO method can be made applicable 
to off-lattice systems by developing some efficient method 
for potential switching.

\section{Conclusion Remarks}
\label{sec:conclusion}
In the present work, we have proposed a new MC method 
based on the stochastic potential switching 
algorithm.~\cite{Mak05,MakSharma07} 
To our knowledge, 
this is the first method for general long-range interacting 
systems that greatly reduces computational time 
without any approximation. This method is 
applicable to any lattice system with long-range interactions. 
The efficiency of the SCO method has been demonstrated 
by applying it to a two-dimensional magnetic dipolar system. 
We have also discussed the properties 
of the SCO method in comparison with those of other methods 
for long-range interacting systems. 

\section*{Acknowledgment}
This work is supported by a Grant-in-Aid for Scientific 
Research (\#18740226) from MEXT in Japan. 

%We have applied the SCO method to a magnetic dipolar systems
%to examine how it works in practice. The obtained results 
%has shown that circular order which originates from long-range 
%dipolar interaction is reproduced by the SCO method 
%with moderate computational time. 

%For instance, in dipolar systems computational time per one MC step 
%is ${\cal O}(N)$ in two dimensions and ${\cal O}(N\log N)$ in three
%dimensions. 

\clearpage
\begin{center}
\large{Figure Captions}\vspace{7mm}
\end{center}
Figure~1 : Set of pairs $\{ {\cal C}_{\rm pair}(\vectr{r}) \}$ 
$(\vectr{r}=(1,1))$ and their labels in a $5\times 5$ square lattice. 
The set consists of $16$ elements. 
\vspace{5mm}\\
Figure~2 : (Color online) Temperature dependences of the specific heat $C$ 
for different sizes. The average is taken over $10$ different runs.
\vspace{5mm}\\
Figure~3 : Snapshot of the spin structure at $T=0.05J$
on a $48\times 48$ square lattice obtained by the SCO method.
\vspace{5mm}\\
Figure~4 : (Color online) Temperature dependences of the circular component of 
the magnetization $M_{\phi}$ for different sizes. 
The average is taken over $10$ different runs. 
\vspace{5mm}\\
Figure~5 : (Color online) Average computational time per MC step 
$t_{\rm av}$ is plotted as a function of $N=L^2$ for the SCO method 
(full squares) and a naive MC method (full circles). 
The solid line and dashed line 
are proportional to $x$ and $x^2$, respectively.
\vspace{5mm}\\
Figure~6 : (Color online) Temperature dependences of the average number 
$\bar{n}$ of potentials $(\bar V_{ij}\ne 0)$ per site.  
The average is taken over $10$ different runs.
\vspace{5mm}\\
Figure~7 : (Color online) Temperature dependences of the specific heat on a 
$48\times 48$ square lattice with different $N_{\rm sw}$ values, 
where $N_{\rm sw}$ is the number of MC steps for every which potentials are 
switched. The data measured in a single run are shown.
\vspace{5mm}\\
Figure~8 : Ratio $t_{\rm switch}/t_{\rm MC}$ is plotted 
as a function of $N=L^2$.
\vspace{5mm}\\
Figure~9 : (Color online) Time evolution of the circular component of 
the magnetization $M_{\phi}$ for $L=48$ when the system is kept 
at $T=0.4J$. The average is taken over $1,000$ different runs. 
The data of the SCO method are denoted by crosses (below) 
and those of the naive MC method are denoted by open squares (above). 
In the inset, $M_{\phi}$ of the SCO method and that of 
the naive MC method are plotted as a function of $t_{\rm s}=t/1.4$ 
and $t$, respectively. 

\clearpage
\clearpage
\clearpage
\clearpage
\clearpage
\clearpage
\clearpage
\clearpage
\clearpage

%\clearpage
%\begin{figure}
%\begin{center}
%\includegraphics[width=12cm]{/home/msasaki/dipole/2D/Dana/cutoff_L48config.eps}
%\end{center}
%\caption{A snapshot of the spin structure at $T=0.05J$ 
%on a $48\times 48$ square lattice obtained by 
%a naive cutoff method with the cutoff distance of $2.7$. }
%\label{fig:snapshot2}
%\end{figure}

\end{document}